\documentstyle[12pt]{article}

\newcommand{\acknowledgements}[1]{\vspace{7mm} \noindent {\normalsize \bf
Acknowledgments.\,} {\normalsize #1}}

\newcommand{\titre}[1]{\begin{center}#1\end{center}}

\include{psfig} \psnoisy 

\begin{document}
 
\baselineskip=24pt

\vspace{1cm}

\titre{\large \bf The optically--dark side of galaxy formation}

\vspace{2cm}

\titre{\bf Bruno Guiderdoni$^1$, Fran\c cois R. Bouchet$^1$, Jean--Loup
Puget$^2$,}

\titre{\bf Guilaine Lagache$^2$ \& Eric Hivon$^3$}

\titre{\it $^1$ Institut d'Astrophysique de Paris, CNRS, 98bis Boulevard
Arago, F--75014 Paris}

\titre{\it $^2$Institut d'Astrophysique Spatiale, B\^at. 121, Universit\'e
Paris XI, F-91405 Orsay Cedex}

\titre{\it $^3$ Theoretical Astrophysics Center, Juliane Maries Vej 30,
DK--2100 Copenhagen}

\vspace{2cm}

{\bf Deep optical surveys \cite{Letal95, Wetal96} probe the rest--frame
ultraviolet luminosities of high--redshift galaxies, which can be converted
into star formation rates under plausible assumptions on young stellar
populations.  The current analysis of these data suggests that the global star
formation rate of the universe peaked at a redshift of 1 and declined since
then \cite{Letal96, Metal96}. This has led to claims that the bulk of star
formation in the universe has been seen. However, the conversion of UV
luminosities into star formation rates must take into account a correction for
the luminosity fraction absorbed by the dust which is generically associated
to young stars. Since this correction is rather uncertain for high--redshift
galaxies, the star formation rates currently deduced from optical surveys
alone might be substantially underestimated. To circumvent this problem, the
simplest is to observe the dust thermal emission at infrared (and
submillimetre) wavelengths and compute the overall luminosity bugdet of
galaxies. For high--redshift galaxies, the only direct observational constraint
is set by the recent detection of the Cosmic Infrared Background (CIRB) built
up from the accumulated IR light of faint galaxies along the line of sight
\cite{Petal96}. Here we propose a more accurate determination of
this long--sought background which solves the main possible weakness of
the earlier determination. Then we estimate the population of high--redshift,
dust--enshrouded starburst galaxies needed to produce this background. We
argue that most of the star formation at high redshift may be hidden by dust,
and we define the necessary characteristics of a feasible
survey at a wavelength of 175 $\mu$m, which could detect this population.} 


While only one third of the bolometric luminosity of local galaxies is
radiated in the IR \cite{SN91}, there is a growing evidence that this fraction
is actually increasing with redshift.  The deepest counts available from the
Infrared Astronomical Satellite (IRAS) at 60 $\mu$m \cite{HH87}, which
correspond to an average redshift $z$ of only 0.2 \cite{AHetal96}, already
suggest some evolution of the IR emission in the universe.  A recent deep
survey with the Infrared Space Observatory (ISO) at 15 $\mu$m has discovered a
few objects at $z \sim 0.5$ to 1 with star formation rates much higher than
deduced from the optical \cite{Retal97}. However, the strongest constraint on
the high--redshift IR emission is given by the CIRB found in data acquired by
the FIRAS instrument on--board the COBE satellite in the 200 $\mu$m -- 2 mm
wavelength range \cite{Petal96}. Several steps are necessary in order to
remove the foregound Galactic components and extract the isotropic residual
identified as the CIRB. While, at the wavelengths probed by FIRAS, the
interplanetary emission is small and easily removed \cite{Retal95}, the
emission from interstellar dust mixed with the different gas phases of the
interstellar medium is the dominant component.  The spectra which correlate
with the 21 cm interstellar emission of neutral hydrogen (HI) \cite{HB95}
along lines--of--sight with HI column densities $N_{HI} \leq 4.5 \times
10^{20}$ atoms cm$^{-2}$ are described by a modified black--body $\nu^2 B_\nu(T)$
with $T=17.5$ K \cite{Betal96}. Part of the long--wavelength excess over this
simple model increases with HI column density and can be linked to dust
emission associated with ionized and molecular hydrogen.  The other part is an
isotropic residual which has been interpreted as the CIRB. The determination
of the isotropy required to use a large fraction of the sky. This thus
involved correcting for substantial foreground components. Any inaccuracy
in this correction might have contributed a spurious signal. In order to
address this problem, the original method of Puget {\it et al.} 
\cite{Petal96} was applied again, but only in the cleanest regions with 
very low HI column densities
($N_{HI} \leq 1 \times 10^{20}$ atoms cm$^{-2}$ instead of
$N_{HI} \leq 4.5 \times 10^{20}$ atoms cm$^{-2}$). In that case, the residual
component totally dominates the emission (inset in fig.1).  This
demonstrates that it cannot be due to artifacts in the removal of interstellar
emission.  Fig.1 also shows that the CIRB intensity per frequency decade $\nu
I_{\nu} = 7~10^{-9}$ W m$^{-2}$ sr$^{-1}$ near 300 $\mu$m is a factor of 5
higher than the no--evolution prediction obtained by a simple extrapolation of
the IR luminosities of local galaxies.  Its level is comparable to that of the
``Cosmic Optical Background'' estimated by summing up faint galaxy counts down
to the deepest limit so far available which is given by the Hubble Deep Field
\cite{Wetal96}.

In order to break the background into the contributing sources, we need to
model the IR/submm emission of galaxies.  Galaxy formation and evolution can
be briefly sketched as follows: Small fluctuations in the high-$z$ universe
grow by gravitational instability until they form dense clumps where the
baryonic gas is collisionally heated. Where the density and temperature are
appropriate, gas cools by emitting radiation, and the baryons pile up in cold
cores whose final radii are set by angular--momentum
conservation. Simultaneously, larger clumps form and encompass/accrete the
previous generation of small clumps. Stars form in the cores and enrich the
primordial gas via supernova ejecta. Part of the starlight is absorbed by
various dust components which re--radiate at longer wavelengths according to
characteristic IR spectra. The so--called ``semi--analytic modelling'' of
these series of physical processes has been rather successful in reproducing
the overall properties of galaxies in the optical range \cite{LS91, KWG93}. We
have elaborated an extension of this method to the IR/submm range. 
Details
of the modelling, and complementary predictions will be given elsewhere
\cite{Getal97, HGB97}.

Specifically, we assume a standard Cold Dark Matter cosmological model
\cite{Betal84} with a Hubble constant $H_0=50$ km s$^{-1}$ Mpc$^{-1}$, a
density parameter $\Omega_0=1$, a cosmological constant $\Lambda=0$, a baryon
fraction $\Omega_b=0.05$, and a normalisation $\sigma_8=0.67$ for the power
spectrum of linear fluctuations. The sensitivity of the semi--analytic 
modelling
to cosmological parameters is known to be weak \cite{Hetal95}. In our
reference scenario A, we consider a mix of two broad types of populations, one
with low star formation rates (which reproduce the observational distribution
of gas consumption time scales in disk galaxies \cite{KTC94}), the other
proceeding in bursts with ten times larger star formation rates. We assume
that bursts are triggered by interaction and merging of sub--galactic clumps
\cite{PWetal96}, and increase with $z$ according to the fraction of pairs
\cite{ZK89, Betal94, CPI94}.  This population of ``mild starbursts'' and
``luminous UV/IR galaxies'' (similar to ``LIRGs'' \cite{SM96}) 
dominate the optical background (fig.1).
The IR luminosity--to--mass ratio is $L_{IR}/M = 6 L_{bol\odot}/M_\odot$ for a
typical starburst with $t_\star=0.33$ Gyr.  The range of derived IR--to--blue
luminosity ratios is characteristic of blue--band selected samples
\cite{Setal87} like the Canada--France Redshift Survey (selected in the
observer--frame $I_{AB}$ band, roughly corresponding to the $B$ band at $z
\sim 1$), or the high--$z$ galaxies of the Hubble Deep Field.  Fig.1 also
displays the predicted IR background, which is clearly barely compatible with
the observed CIRB whose mean amplitude is twice higher.
 
In order to assess how much star formation might be {\it completely} hidden by
dust shrouds, we consider an additional population, similar to
``ultra-luminous IR galaxies'' (``ULIRGs'') \cite{SM96}. We maximize their IR
luminosities by assuming that all the energy available from stellar
nucleosynthesis is radiated by massive stars and heats up the dust.  As a
consequence of a stellar initial mass function with short--lived, massive
stars, the post--starburst phase is ``dark'' and would be detectable only by
its nucleosynthesis products.  The luminosity--to--mass ratio now is $L_{IR}/M
= 130 L_{bol\odot}/M_\odot$ for a typical starburst with $t_\star=0.33$
Gyr. Our scenarios B and C respectively mimic continous bulge formation as the
end--product of interaction and merging, and a strong episode of bulge
formation at $z_{for} > 3.5$. Both are consistent with the observed CIRB.  The
high--redshift, dust--enshrouded, star formation in scenario C results in the
high level of the predicted background at wavelengths $\lambda > 400$ $\mu$m.

While none of the currently available optical data reflects the
large differences between these scenarios, originating in the different
fractions of heavily--extinguished objects, the predicted IR/submm counts 
are more interesting, as shown in fig.2.  The comparison of IRAS
data with the no--evolution curve at 60 $\mu$m suggests some evolution.
However, it appears that the 60 $\mu$m band does not strongly discriminate
between the various scenarios of evolution. In contrast, the upward deviation
at 200 $\mu$m is due to the contribution of the redshifted 100 $\mu$m maximum
of the IR energy distribution. This redshifting of steep spectra
counter--balances distance dimming and can make high--$z$ objects {\it
easier} to detect than low--$z$ ones. Submm observations are thus quite
sensitive to the high--$z$ history. The model also predicts that, at 200
$\mu$m, 10--100 mJy sources (contributing to 15 \% of the background) are
mostly located at $z \sim 0.5$ -- 2.5, while at 60 $\mu$m, and at the typical
sensivity level of IRAS surveys, the sources are indeed located mostly at very
low $z$.

The detection of these sources would be a strong test for assessing the level
of the ``optically--dark'' side of galaxy formation.  The C160 filter of the
ISOPHOT instrument on--board ISO has an effective wavelength $\lambda_{eff}
\simeq 175$ $\mu$m for typical spectra of distant galaxies, and a 10 mJy {\it
rms} noise fluctuation per 1.5 arcmin pixel is reachable after integration times
larger than $\sim 256$ s per pixel. Thus a deep survey with this instrument
appears to be feasible and is indeed scheduled.  However, one might be
concerned that small--scale cirrus fluctuations could hide the sources and the
fluctuations of the background they induce. A comparative analysis of the
expected power spectra due to (1) cirrus fluctuations in regions of various HI
column densities, (2) background fluctuations once sources above the confusion
limit have been removed, (3) the detector noise, shows that, in clean regions
of the sky ($N_{HI} \leq 1\times 10^{20}$ atoms cm$^{-2}$), a survey with 10
mJy {\it rms} sensitivity should not only detect most sources above the low
cirrus fluctuations but could also see background fluctuations in excess of
the detector noise fluctuations, on scales 3 to 10 arcmin (see fig.3).  Since
scenario C has $6.3 \times 10^5$ sources/sr with fluxes $> 30$ mJy, a deep
survey of a $\sim$ 1000 arcmin$^2$ field might begin to ``break'' the CIRB
into $\sim 50 \pm 7$ discrete sources, an order of magnitude more than is
expected without evolution. This number is sufficient to test the high level
of evolution and even begin to disentangle between the various scenarios.  If
detected, the level of background fluctuations can also help to constrain
the redshift distribution of the sources.

The tentative discovery of the Cosmic Infrared Background is now supported by
our new study in the cleanest regions of the sky, where the foregound Galactic
components are essentially negligible. The conversion of the CIRB into its
contributing sources by means of a semi--analytic model of galaxy formation
leads to predictions of faint galaxy counts at IR and submm wavelengths.  In
contrast with the status of IRAS 60 $\mu$m counts, 175 $\mu$m counts with ISO
should be able to detect the predicted strong evolution, and even disentangle
between various scenarios consistent with the level of the CIRB. Moreover,
small--scale cirrus fluctuations cannot hide the presence of the sources.
Consequently, our knowledge of the optical/IR luminosity budget at $z \sim
0.5$ -- 2.5 should improve rapidly. We are about to start unveiling the
optically--dark side of galaxy formation.


 \acknowledgements{ We are pleased to thank Dave Clements, Fran\c cois--Xavier
D\'esert, and Bruno Maffei for their comments and suggestions.}

\pagebreak


\centerline{\bf Figure Captions}

\bigskip
\bigskip
{\bf Figure 1:} {\it Inset panel:} High--latitude COBE/FIRAS spectrum in
regions with very low HI column densities ($N_{HI} \leq 1 \times 10^{20}$
atoms cm$^{-2}$, solid line) and residual spectrum after subtraction of emissions
correlated with neutral and ionized hydrogen (dotted line).  This component is
likely to be the Cosmic Infrared Background (CIRB).  {\it Main panel:} The
$\pm 1\sigma$ error bars per point have been used to define an acceptable
range for CIRB predictions (thick solid lines). Solid squares show the upper
limits given by COBE/DIRBE residuals \cite{H96}.  The solid hexagons show the
Cosmic Optical Background obtained by summing up faint galaxy counts down to
the Hubble Deep Field limit. Strictly speaking, this is only a lower limit of
the actual optical background, but the shallowing of the $U$ and $B$--band
counts suggests near--convergence at least at those wavelengths
\cite{Wetal96}.  The short and long dashes show the background predicted
without evolution, in a universe with $H_0=50$ km s$^{-1}$ Mpc$^{-1}$ and
$\Omega_0=1$.  The solid, dotted and dashed lines show the predictions for our
scenarios A, B, and C respectively, in the standard CDM model
($H_0=50$ km s$^{-1}$ Mpc$^{-1}$, $\Omega_0=1$, $\Lambda=0$, $\Omega_b=0.05$,
and $\sigma_8=0.67$).  Stars form in cold cores according to Salpeter's
stellar initial mass function (with slope $s=1.35$, and lower and upper mass
cut--offs 0.1 and 120 $M_\odot$). Star formation rates are proportional to the
cold gas contents, with characteristic time scales derived from the core
dynamical times as $t_\star = \beta t_{dyn}$. Scenario A has a mix of two
populations, a disk--like one with $\beta=100$, and starbursts with
$\beta=10$.  The mass fraction involved in bursts increases with the formation
redshift $z_{for}$ according to $f_{burst} \propto (1+z_{for})^5$, as
suggested by the increasing fraction of pairs seen at larger $z$ \cite{ZK89,
Betal94, CPI94}. In scenario B and C, we add a population of ``ultra-luminous
IR galaxies'' (ULIRGs) in which all the energy available from stellar
nucleosynthesis ($0.007 x M c^2$) is radiated by massive stars ($<x>=0.40$) in
a heavily--extinguished medium.  Scenario B has a 5 \% constant mass fraction
of ULIRGs at all $z_{for}$.  In scenario C, 90 \% of all galaxies forming at
$z_{for} \geq 3.5$ are ULIRGs.

\bigskip
{\bf Figure 2:} Predictions for differential counts normalized to Euclidean
counts at 60 $\mu$m (upper panel) and 200 $\mu$m (lower panel).  The short and
long dashes show the predicted counts without evolution, in a universe with
$H_0=50$ km s$^{-1}$ Mpc$^{-1}$ and $\Omega_0=1$.  The solid, dotted and
dashed lines show the predictions for our scenarios A, B, and C respectively
in the SCDM model (see text). Data are plotted for IRAS counts at 60
$\mu$m. Open stars: Faint Source Survey \cite{Letal90}. Open squares: QMW
survey \cite{Retal91}.  Solid squares: North Ecliptic Pole Region \cite{HH87}.
The IRAS 60 $\mu$m counts suggest some evolution, but do not discriminate
between the various scenarios. These scenarios also predict a 60 $\mu$m
background fluctuation per beam in the Very Faint Source Survey (after removal
of $\geq 4\sigma_{tot}=120$ mJy sources) at the level of 14.1 mJy (A), 16.0
mJy (B), and 14.3 mJy (C), while the measured 68 \% quantile is $30.1 \pm$ 1.2
mJy \cite{BDM96}.  With a 25 mJy {\it rms} instrumental noise and 6.5 mJy {\it
rms} cirrus fluctuations \cite{Getal92}, there is still space for a
$15.4^{+2.2}_{-2.5}$ mJy fluctuation due to sources, in good agreement with
our estimates.  In contrast, scenarios A, B, C
predict much stronger evolution at
200 $\mu$m.  The differences in the predicted counts originate from the
differences in the high--redshift IR emissions.

\bigskip
{\bf Figure 3:} Comparison of predicted power spectra for observations in the
ISO C160 filter. The straight horizontal lines (same code as in fig.1) show 
the predicted background fluctuations at the confusion limit (i.e. its {\it
rms} value equals one third of the limiting flux of all resolved and removed
sources). The beam was modelled as a 1 arcmin FWHM gaussian. 
The cirrus fluctuations
(thin lines) are described by a $k^{-3}$ power law \cite{Getal92} with
levels (from left to right) set by column densities of neutral hydrogen (HI)
respectively typical of the Lockman hole (minimum, 0.24 \% of the sky), and of
4.6 \%, \& 21.8 \% of the sky. The detector noise spectrum of
fluctuations (thin dashed lines), assumed to be white, is shown for a
sensitivity level of 10 and 20 mJy per 1.5 arcmin pixel. To allow for a
direct comparison, this flat white noise has been projected on the sky,
i.e. divided by the Fourier transform of the beam profile. The resulting
exponential rise of the on--sky noise thus bounds the range of available
scales. The comparison shows that, in the cleanest regions of the sky,
(e.g. $N_{HI} = 1 \times 10^{20}$ atoms cm$^{-2}$), 
unresolved background fluctuations for
scenarios B and C actually dominate both cirrus and noise fluctuations at
scales $3 < \theta < 10$ arcmin, if the {\it rms} noise level is 10 mJy per
pixel.

\begin{figure}[htbp] 
\psfig{figure=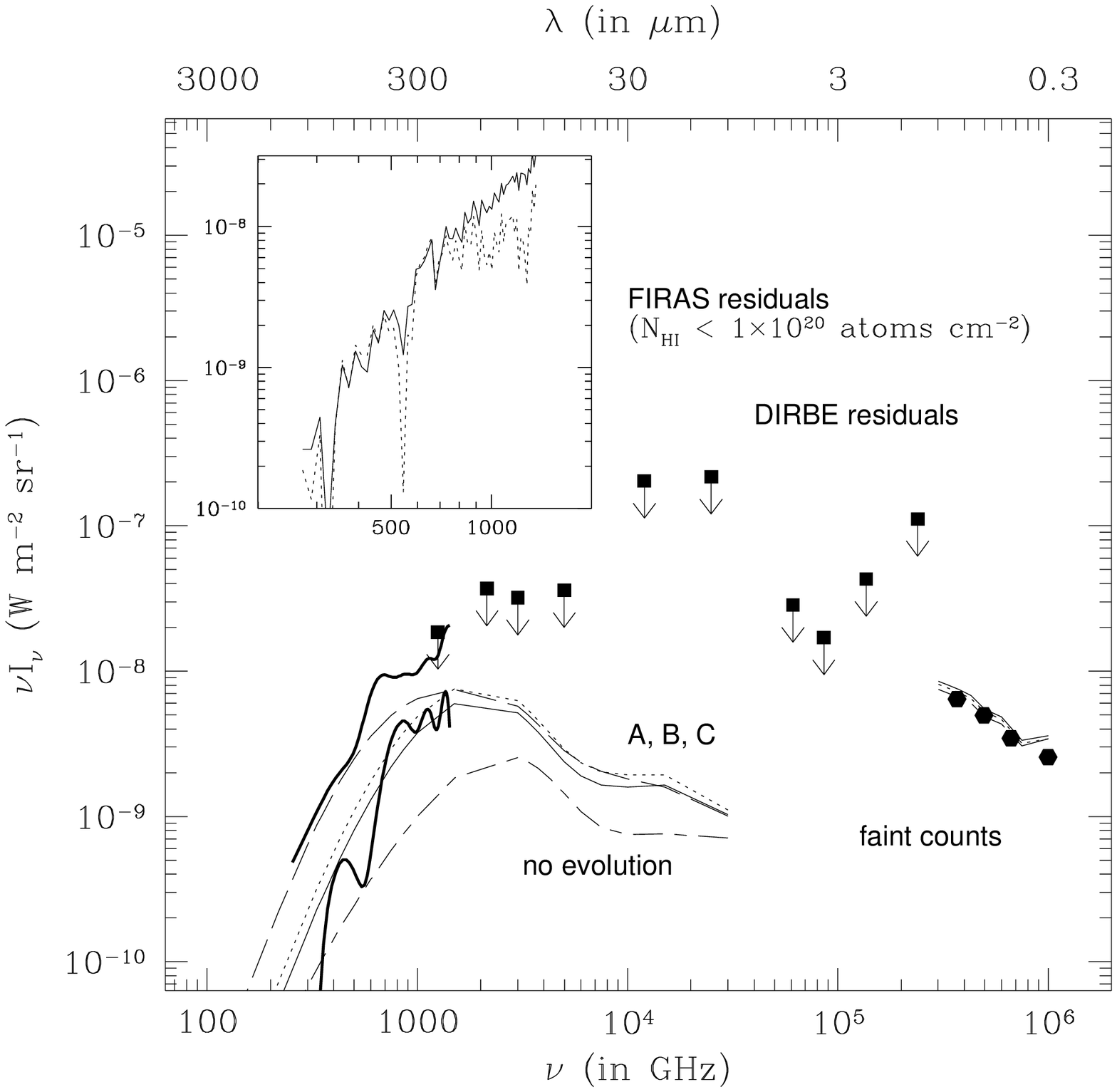,width=\textwidth}
\caption[ ]{ }
\end{figure}

\begin{figure}[htbp]
\psfig{figure=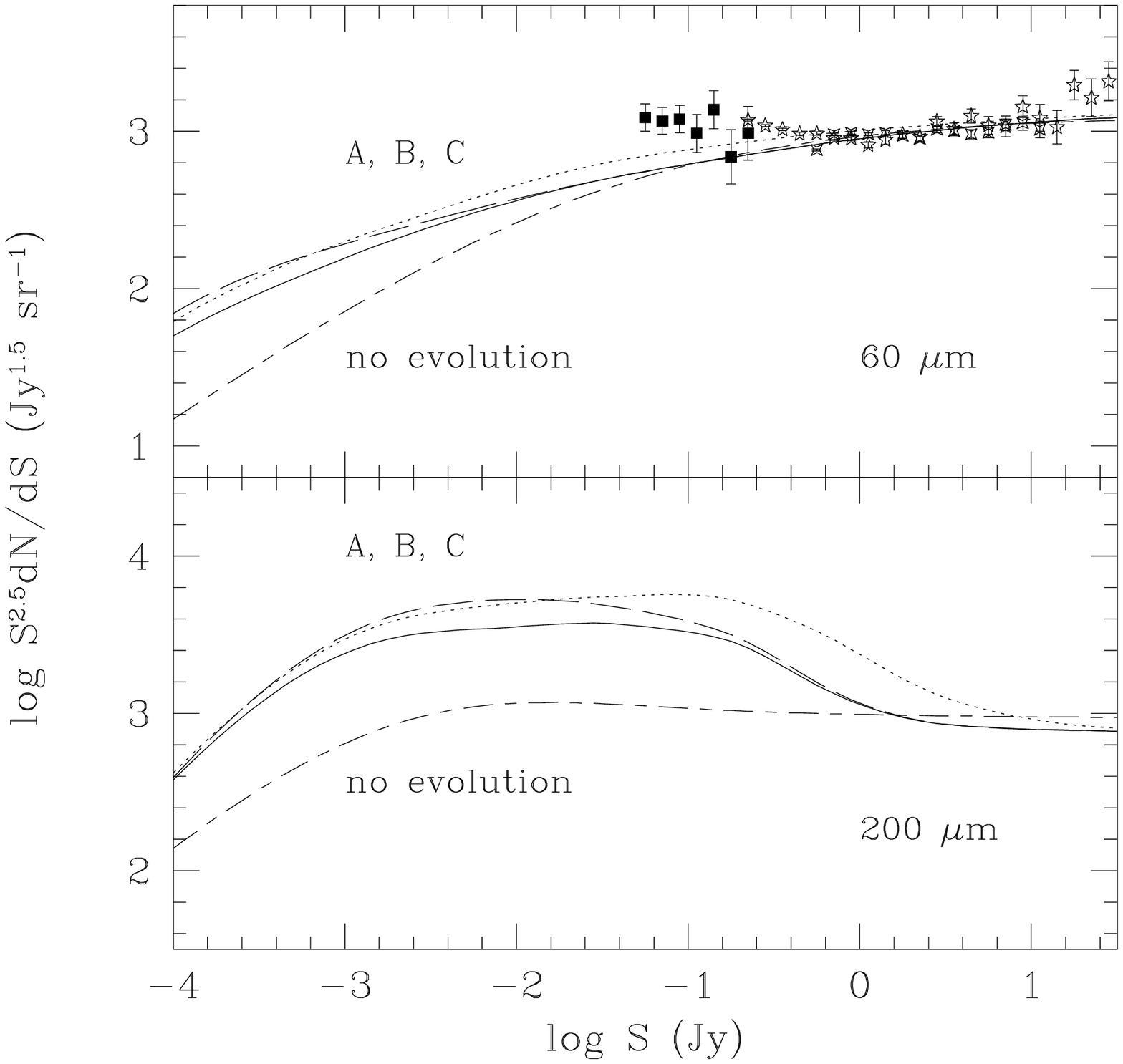,width=\textwidth}
\caption[ ]{ }
\end{figure}

\begin{figure}[htbp] 
\psfig{figure=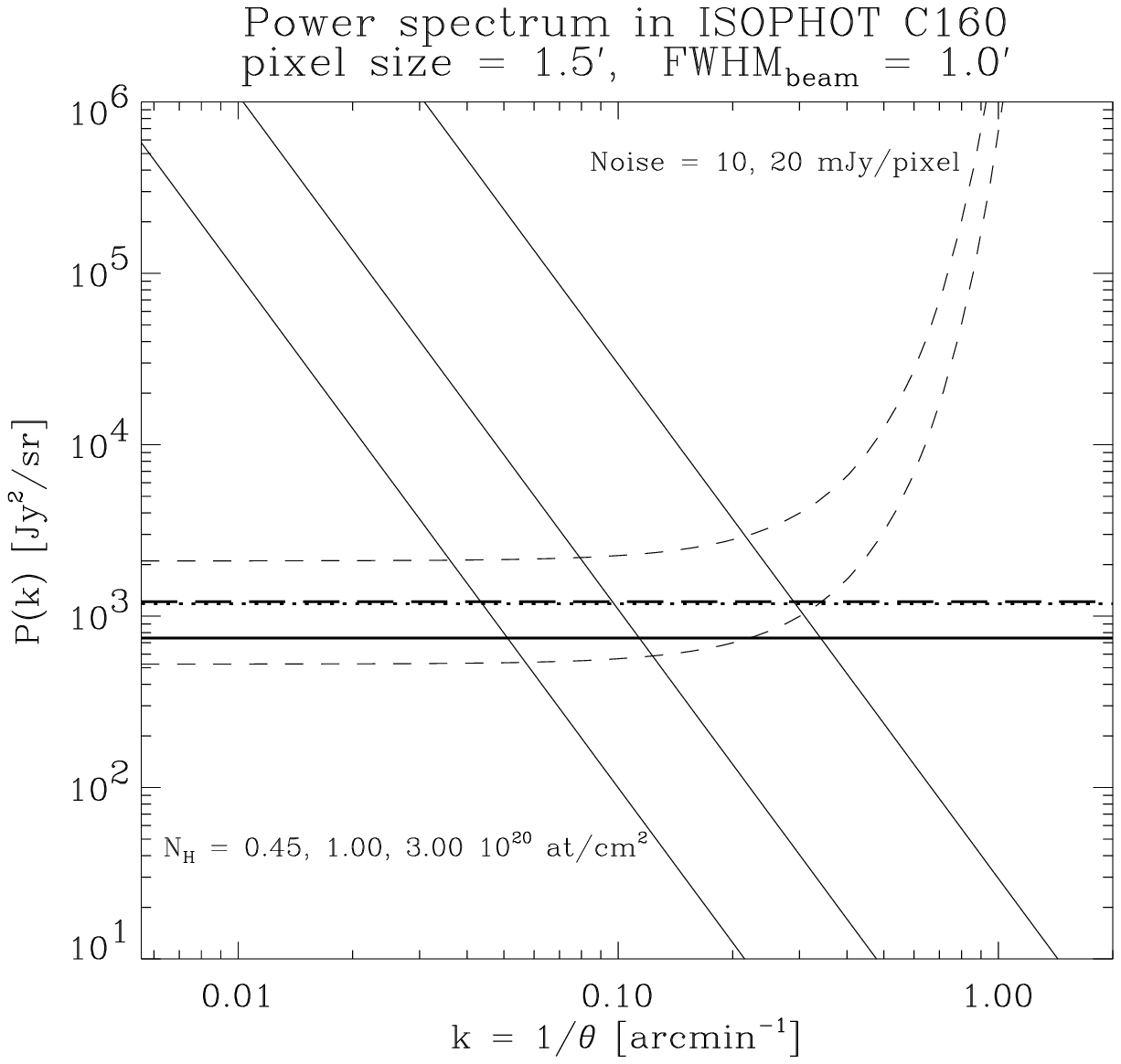,width=\textwidth}
\caption[ ]{ }
\end{figure}

\end{document}